\documentclass[a4paper,11pt,twoside,openright]{article}

\usepackage[francais]{babel}                                    
\usepackage[dvips,pdftex]{graphicx}
\usepackage[left=2.5cm,top=3cm,right=2.5cm,bottom=3cm]{geometry}    
\usepackage[latin1]{inputenc}                                   
\usepackage{amsmath,amsthm,amssymb}                             
\usepackage{fancyhdr}
\usepackage{color}

\title{Titan's Obliquity as evidence for a subsurface ocean?}
\author{Rose-Marie Baland , Tim Van Hoolst, Marie Yseboodt and \"{O}. Karatekin\\
\textit{Royal Observatory of Belgium, Brussels, Belgium}\\Email: balandrm@oma.be}
\date{April 2011\\ The final version of this preprint is published in Astronomy \& Astrophysics \\DOI: 10.1051/0004-6361/201116578 }

\begin{document}

   \maketitle
 
\section*{Abstract}

On the basis of gravity and radar observations with the Cassini spacecraft, the moment of inertia of Titan and the orientation of Titan's rotation axis have been estimated in recent studies. According to the observed orientation, Titan is close to the Cassini state. However, the observed obliquity is inconsistent with the estimate of the moment of inertia for an entirely solid Titan occupying the Cassini state. We propose a new Cassini state model for Titan in which we assume the presence of a liquid water ocean beneath an ice shell and consider the gravitational and pressure torques arising between the different layers of the satellite. With the new model, we find a closer agreement between the moment of inertia and the rotation state than for the solid case, strengthening the possibility that Titan has a subsurface ocean.

\section{Introduction}
 
On the basis of Cassini radar images, \cite{bwstiles} and \cite{bwstiles2} precisely measured  the orientation of the rotation axis of Titan. Using the orientation of the normal to the orbit of Titan given in the IAU recommendations (Seidelmann et al. 2007), they determined the obliquity to be about $0.3 ^\circ$. They also showed that the rotation axis makes a small angle of about $0.1^\circ$ with respect to the plane defined by the normal to the Laplace plane and the normal to the orbit and, as a result, they stated that Titan is close to the Cassini state. \\
\indent From Cassini radio tracking, the quadrupole field of Titan has been found to be consistent with a body in hydrostatic equilibrium with a moment of inertia $C/M R^{2}=0.3414\pm0.0005$ \cite{iess}. However, in a study of the Cassini state generalized to a multi-frequency orbital node precession in which Titan was considered as an entirely solid body, \cite{billsepsc09} found that the $0.3^\circ$ obliquity implies a moment of inertia of $C=0.55M R^{2}$, which is not only inconsistent with the above hydrostatic value of the moment of inertia but would also imply a physically implausible interior structure with higher mass density towards the surface than towards the center. In their conclusion, the authors proposed that a liquid ocean could partially decouple the shell from the interior.\\
\indent The obliquity of the Cassini sate, which is an equilibrium orientation of the rotation axis of a synchronous satellite, can be derived from angular momentum equations in much the same way as for the periodic variations about its equilibrium rotation rate, but with different assumptions concerning the timescales involved (long timescales for the obliquity and short timescales for the length-of day variations). The influence of a global liquid layer on the LOD variations of Titan and on the librations of the Galilean satellites was investigated in \cite{vanhoolst} and \cite{baland}, respectively. Here, we extend their method, considering the appropriate timescales, to the Cassini state. By considering the gravitational and pressure torques between the different layers of the satellite, we show that the orientation of the rotation axis given in \cite{bwstiles} can be partially reconciled with the moment of inertia given in \cite{iess}.

\section{The Cassini state for a solid Titan}

\begin{table*}
\caption{\textbf{Orbital theory of Titan and solid Cassini state:} columns 2 to 5: Amplitudes, frequencies, periods, and phases of the orbit precession adapted from \cite{vienne}. They give the orbital precession in the equatorial plane of Saturn and used J1980 as the time origin. Here we consider the Laplace plane and J2000 time origin. The Laplace plane has a node of $184.578^\circ$ and an inclination (also called tilt) of $0.6420^\circ$ with respect to the equatorial plane of Saturn. Since the tilt is small, the frequency and amplitude of the orbital precession are almost the same with respect to the Laplace plane as to the equatorial plane of Saturn. The x-axis of the Laplace place is taken as the node of the Laplace plane on the equatorial plane of Saturn. The obliquity amplitudes and resonance factors  ($fr_j$) of the solid Cassini state model presented in Section 3 are given in the last two columns.}             
\label{table1}      
\centering          
\begin{tabular}{c c c c c c c }   
\hline\hline   $j$&$i_j$&$\dot\Omega_j$&period&$\gamma_j$&$\varepsilon_j$ &$fr_j$  \\
  &(deg)&(rad/year)& (years)&(deg)&(deg)\\    
\hline                    
      $1$&$0.3197$&$-0.00893124$&$-703.51$&$160.691$&$0.1199$&$1.38$   \\
      $2$&$0.0150$&$-0.00192554$&$-3263.07$&$102.230$&$0.0009$&$1.06$ \\
      $3$&$0.0129$&$0.42659824$ &$14.73$&$292.867$&$-0.0120$&$-$ \\
      $4$&$0.0022$&$-0.21329912$&$-29.46$&$222.920$&$-0.0026$&$1.18$ \\
\hline
\end{tabular}
\end{table*}

In the classical Cassini state, the rotation axis, the normal to the orbit, and the normal to the Laplace plane of a synchronous solid satellite remain in the same plane since the rotation axis has the same constant precession rate as the normal to the orbit about the normal to the Laplace plane, which is, by definition, the mean orbital plane of the satellite. The obliquity is then the constant angle between the rotation axis and the normal to the orbit.

To be able to develop the Cassini state in the presence of a liquid subsurface ocean, we first present the generalization of the Cassini state of a solid Titan to a multi-frequency node precession, following \cite{bills05}. Neglecting wobble, the rotation axis coincides with the principal axis of the polar moment of inertia $C$ and the angular momentum equation is
\begin{eqnarray}
  n_T C \frac{d\hat s}{dt}&=&n_T \kappa (\hat s \wedge \hat n),\nonumber \\
\label{rigidcase}\kappa&=&\frac{3}{2}MR^2(-C_{20}+2C_{22})n_T=\frac{3}{2}(C-A)n_T,
\end{eqnarray}
where $\hat s=(s_x,s_y,s_z)$ and $\hat n=(n_x,n_y,n_z)$ are the unit vectors along the rotation axis and the normal to the orbit, expressed in coordinates $(x,y)$ of the Laplace plane and $z$ along the normal. In addition, $M$ and $R$ are the mass and mean radius, $C_{20}$ and $C_{22}$ are the second-degree gravity field coefficients, $A$ is the smallest moment of inertia, and $n_T$ is the mean motion, equal to the rotation rate, of Titan. The right-hand member is the gravitational torque (averaged over the orbital period) exerted by Saturn. The vectorial equation is then projected on the Laplace plane, where the motions of the projected rotation axis and orbit normal are easy to parametrize
\begin{eqnarray}\label{rigidprojected}
 \frac{d S}{dt}=I \frac{\kappa}{C} (N-S)
\end{eqnarray}
with $S=s_x+I s_y$ and $N=n_x+I n_y$, $I=\sqrt{-1}$. Equation (\ref{rigidprojected}) is correct up to the first order in small obliquity, eccentricity and inclination (see Eq. (46) of \cite{bills05} in which a sign has been corrected). We first assume that the orbital precession $N$, which causes the rotation axis precession, is zero in order to get the free spin precession mode whose frequency $\omega_f$ depends only on the physical properties of the satellite
\begin{eqnarray}
 \omega_f=\frac{\kappa}{C}.
\end{eqnarray}
To compute the forced solution, the orbital precession is written as a series expansion 
\begin{eqnarray}\label{precession}
N=\sum_j\sin{i_j}\, e^{I (\dot\Omega_j t+ \gamma_j-\pi/2)},
\end{eqnarray}
where the inclination amplitudes $i_j$, frequencies $\dot\Omega_j$, and phases $\gamma_j$ associated with the node precession of the orbit with respect to the Laplace plane have been taken, up to $j=4$, from \cite{vienne} and are given in Table 1. The parameters $\dot\Omega_1$ and $i_1$ are the main precession rate (with a period of 703.51 years) and the small mean inclination ($0.3197^\circ$) of the orbital plane, considered in the classical Cassini state. The forced solution of Eq.(\ref{rigidprojected}) is then
\begin{eqnarray}
S&=&\sum_j \sin{(i_j+\varepsilon_j)}\, e^{I (\dot\Omega_j t+ \gamma_j-\pi/2)},
\end{eqnarray}
where, correct up to the first order in $\varepsilon_j$ and $i_j$, the obliquity amplitude $\varepsilon_j$ associated with the frequency $\dot\Omega_j$ is given by
\begin{eqnarray}\label{rigidfirstorder}
\varepsilon_j=-\frac{i_j\dot\Omega_j}{(\omega_f+\dot\Omega_j)}.
\end{eqnarray}
The obliquity $\varepsilon$ at any time is the non-constant angle between the rotation axis and the normal to the orbit and oscillates between two extreme values $\varepsilon_{min}$ and $\varepsilon_{max}$ such that
\begin{eqnarray}\label{obl}
\varepsilon&\cong&\sin{\varepsilon}=\|S-N\|\\
\label{minmax}
\varepsilon_{min}&=&2\, \max_j\{|\varepsilon_j|\}-\sum_{j} |\varepsilon_j|\leq \varepsilon\leq \sum_j |\varepsilon_j|=\varepsilon_{max}.
\end{eqnarray}
In this generalization, the normal to the Laplace plane, the normal to the orbit, and the rotation axis are not coplanar and the small angular deviation $\delta$ of the spin axis with respect to the plane defined by the other axes is
\begin{eqnarray}\label{dev}
\delta&\cong& \sin{\delta}=(n_x s_y- n_y s_x)/\|N\|.
\end{eqnarray}
Since the frequency $\omega_f$ is positive, $\omega_f+\dot\Omega_j$ can be close to zero if $\dot\Omega_j <0$, and the coefficient $\varepsilon_j$ can be amplified by a resonance. For $\dot\Omega_j <0$, we define \textquotedblleft resonance factors\textquotedblright, $fr_j$, which describe the amplification of $\varepsilon_{j}$ caused by the resonance and are close to one far from resonance, given by
\begin{eqnarray}\label{fr}
fr_j=\max(|\omega_f|,|\dot\Omega_j|)/|\omega_f+\dot\Omega_j|
\end{eqnarray}

For $C_{20}=-31.808\times10^{-6}$, $C_{22}=9.983\times10^{-6}$, and $C=0.3414 MR^2$ (Iess et al. 2010), the resonant factors $fr_j$ are close to one  because the free mode period is $191.91$ years. Therefore, no significant resonant amplification (see Table 1) occurs and the obliquity variations are small. The obliquity $\varepsilon$ of about $0.12^\circ$, and the maximal deviation $\delta$ of about $0.02^\circ$ (Fig. 4) are inconsistent with the results of \cite{bwstiles}.

\section{Titan with a liquid ocean}

\subsection{A new Cassini state solution}

We now assume that Titan consists of four homogeneous layers: an ice shell ($sh$), a liquid ocean ($o$), an ice mantle ($m$), and an ice/rock core ($c$). The solid layer composed of the mantle and the core is also called interior ($in$) hereafter. The shell and the interior are considered to behave rigidly. For the solid layers, the angular momentum vector $\vec H_l$ can be written as the product of the polar moment of inertia $C_l$ and the rotation vector $n_T\hat s_l$. The angular momentum equations take the form
\begin{eqnarray}
 \label{12a} n_T C_{sh} \frac{d\hat s_{sh}}{dt}&=&\vec \Gamma_{sh,ext}+\vec \Gamma^p_{sh,ext}+\vec \Gamma_{sh,int}+\vec \Gamma^p_{sh,int}, \\
 \label{12b}\frac{d \vec H_o}{dt}&=&\vec \Gamma_{o,ext}+\vec \Gamma^p_{o,ext}+\vec \Gamma_{o,int}+\vec \Gamma^p_{o,int}, \\
 \label{12c} n_T C_{in} \frac{d\hat s_{in}}{dt}&=&\vec \Gamma_{in,ext}+\vec \Gamma^p_{in,ext}+\vec \Gamma_{in,int}+\vec \Gamma^p_{in,int},
\end{eqnarray}
where the external gravitational torque exerted by Saturn on layer ($l$), $\vec \Gamma_{l,ext}$, is equal to $-\int_{V_l}(\vec r\wedge \rho_j \nabla W) \,dV$, where $W$ is the gravitational potential of Saturn averaged over the orbital period, $\vec r$ is the position vector of the points located inside the volume $V_l$ of layer ($l$), and $\vec \Gamma_{l,int}$ is the internal gravitational torque due to layers with a different orientation than layer ($l$). The liquid ocean induces pressure torques at the interfaces between the solid layers and the ocean. As we shall see later, the pressure torques can be interpreted as a modification of the external and internal gravitational torques and are therefore denoted by $\vec \Gamma^p_{l,ext}$ and $\vec \Gamma^p_{l,int}$.

The external gravitational torques on the shell and the interior are written as in Eq.(\ref{rigidcase}) for the solid case
\begin{eqnarray}
\vec \Gamma_{sh,ext}=n_T \kappa_{sh} (\hat s_{sh} \wedge \hat n),\, \textrm{with}\, \kappa_{sh}=\frac{3}{2}(C_{sh}-A_{sh})n_T,\\
\label{gammaext} \vec \Gamma_{in,ext}=n_T \kappa_{in} (\hat s_{in} \wedge \hat n),\, \textrm{with}\, \kappa_{in}=\frac{3}{2}(C_{in}-A_{in})n_T.
\end{eqnarray}
We divide the ocean into a top part and a bottom part, respectively, above and below an arbitrary chosen sphere inside the ocean. The pole axes of the top and bottom parts are those of the shell and the interior, respectively, and 
\begin{eqnarray}
\vec \Gamma_{o,ext}&=&n_T \kappa_{o,t}(\hat s_{sh} \wedge \hat n)+n_T \kappa_{o,b}(\hat s_{in} \wedge \hat n),\nonumber\\
\label{oceantorque}\kappa_{o,t}&=&\frac{3}{2}(C_{o,t}-A_{o,t})n_T,\, \kappa_{o,b}=\frac{3}{2}(C_{o,b}-A_{o,b})n_T,
\end{eqnarray}
where $(C_{o,t}-A_{o,t})$ and $(C_{o,b}-A_{o,b})$ are the moment of inertia difference of the top and bottom part, respectively. For the long timescales considered here, it is a very good approximation to assume that the fluid is in hydrostatic equilibrium, therefore $W$ induces a pressure $P_{ext}$ in the ocean such that $\nabla P_{ext}= -\rho_o \nabla W$. The modification of the external torque on the interior because of the pressure is, applying Gauss' theorem, 
\begin{eqnarray}
\label{extpresstorque} \vec \Gamma^p_{in,ext}&=&
\int_{V_{in}}(\vec r\wedge \rho_o \nabla W) \,dV,
\end{eqnarray}
where $\vec r$ is the position vector of the points located inside the volume $V_{in}$ of the interior. It then follows that 
\begin{eqnarray}
 \vec \Gamma^p_{in,ext}&=&n_T \kappa_{o,b} (\hat s_{in} \wedge \hat n).
\end{eqnarray}
In the same way, the modification of the external torques on both the shell and the ocean because of the pressure effect are
\begin{eqnarray}
 \vec \Gamma^p_{sh,ext}&=&n_T \kappa_{o,t} (\hat s_{sh} \wedge \hat n),\\
 \vec \Gamma^p_{o,ext}&=&-n_T \kappa_{o,b} (\hat s_{in} \wedge \hat n)-n_T \kappa_{o,t} (\hat s_{sh} \wedge \hat n).
\end{eqnarray}
By using Eq. (\ref{oceantorque}), we see that $\vec \Gamma_{o,ext}+ \vec \Gamma^p_{o,ext}=0$, which is a consequence of hydrostatic equilibrium.

Two layers with different obliquities and coincident axes of the moment of inertia B (on average, as a consequence of the Cassini state) 
exert a gravitational torque on each other tending to align their pole axis with each other. This internal gravitational torque on any layer ($l$) is given by
\begin{eqnarray}\label{inttorque}
 \vec \Gamma_{l,int}=-\int_{V_l}(\vec r \wedge \rho_l \nabla \Phi) \, dV,
\end{eqnarray} 
where $\Phi$ is the internal gravitational potential, averaged over the orbital period, of the layers with a different obliquity than layer ($l$). From \cite{zseto}, we express the torque on the interior due to the shell and the top ocean and the torque on the shell caused by the misalignment with the bottom ocean and the interior as 
\begin{eqnarray}
 \vec \Gamma_{in,int}&=&-(8\pi G/5)(C_{in}-A_{in}) \left[\rho_{sh}(\alpha_{sh}-\alpha_o+\beta_{sh}/2\right.\nonumber\\
&&\left.-\beta_o/2)+\rho_o (\alpha_o+\beta_o/2)\right] (\hat s_{sh} \wedge \hat s_{in}), \\
\vec  \Gamma_{s,int}&=&(8\pi G/5)[(C_{in}-A_{in})+(C_{o,b}-A_{o,b})]\times\nonumber\\
\label{graviint}&&[\rho_{sh}(\alpha_{sh}-\alpha_o+\beta_{sh}/2-\beta_o/2)] (\hat s_{sh} \wedge \hat s_{in}),
\end{eqnarray}
where $\alpha$ and $\beta$ are the polar and equatorial flattenings, defined as the relative differences $((a+b)/2-c)/(a+b)/2)$ and $(a-b)/a$, respectively, with $a>b>c$ the radii in the direction of the principal axes of the layers. For the ocean, we sum the torques on the top and bottom parts
\begin{eqnarray}
 \vec \Gamma_{o,int}&=&(8\pi G/5)(C_{in}-A_{in})[\rho_o(\alpha_o+\beta_o/2)] (\hat s_{sh} \wedge \hat s_{in})\nonumber\\
&&-(8\pi G/5)(C_{o,b}-A_{o,b})\times\nonumber\\
&& [\rho_{sh}(\alpha_{sh}-\alpha_o+\beta_{sh}/2-\beta_o/2)] (\hat s_{sh} \wedge \hat s_{in}).
\end{eqnarray}
The potential $\Phi$ induces a pressure $P_{int}$ in the ocean ($\nabla P_{int}= -\rho_o \nabla \phi$) that modifies the internal gravitational torque on the interior 
\begin{eqnarray}\label{intpresstorque} 
 \vec \Gamma^p_{in,int}&=&\int_{V_{in}}(\vec r\wedge \rho_o \nabla \Phi) \,dV.
\end{eqnarray}

Because of the similarity between Eq.(\ref{intpresstorque}) and Eq.(\ref{inttorque}) and by using Eq.(\ref{graviint}), we have 
\begin{eqnarray}
 \vec \Gamma^p_{in,int}&=&-(8\pi G/5)(C_{o,b}-A_{o,b})[\rho_{sh}(\alpha_{sh}-\alpha_o+\beta_{sh}/2\nonumber\\
&&-\beta_o/2)+\rho_o (\alpha_o+\beta_o/2)] (\hat s_{sh} \wedge \hat s_{in}).
\end{eqnarray}
In the same way as for the interior, we have
\begin{eqnarray}
\vec \Gamma^p_{sh,int}&=&(8\pi G/5)[(C_{in}-A_{in})+(C_{o,b}-A_{o,b})]\nonumber\\
&&\times[\rho_o(\alpha_o+\beta_o/2)] (\hat s_{sh} \wedge \hat s_{in})
\end{eqnarray}
for the shell.

Therefore, $ \vec \Gamma_{sh,int}+\vec \Gamma^p_{sh,int}+\vec \Gamma_{in,int}+\vec \Gamma^p_{in,int}=0$. Since the sum of all internal torques must be zero, we have that $\vec \Gamma_{o,int}+ \vec \Gamma^p_{o,int}=0$ and that the total torque on the ocean is zero. Therefore, we only need to consider the angular momentum equations for the shell and the interior, given by Eqs.(\ref{12a}) and (\ref{12c}). By introducing
\begin{eqnarray}
 \kappa'_{sh}&=&\kappa_{sh}+\kappa_{o,t},\\
 \kappa'_{in}&=&\kappa_{in}+\kappa_{o,b},\\
 K&=&-[(8\pi G)/(5n_T)] [(C_{in}-A_{in})+(C_{o,b}-A_{o,b})]\times\nonumber\\
 && [\rho_{sh}(\alpha_{sh}-\alpha_o+\beta_{sh}/2-\beta_o/2)+\rho_o (\alpha_o+\beta_o/2)],
\end{eqnarray}
the system of angular momentum equation for the shell and the interior, projected onto the Laplace plane, is

\begin{eqnarray}
 C_{sh} \frac{dS_{sh}}{dt}&=&I \kappa'_{sh}(N-S_{sh})-I K (S_{in}-S_{sh}),\\
 C_{in} \label{system}\frac{dS_{in}}{dt}&=&I \kappa'_{in}(N-S_{in})+I K(S_{in}-S_{sh}).
\end{eqnarray}
The two free modes of the system correspond to 'coupled' (with the frequency $\omega_{+}$) and 'decoupled' ($\omega_{-}$) modes in which the shell and the interior oscillate in the same and opposite directions, respectively, such that
\begin{eqnarray}
\omega_{\pm}&=&-(Z \pm \sqrt{\Delta})/(2 C_{in} C_{sh}),\\
Z&=&K (C_{in}+C_{sh})-C_{sh} \kappa'_{in} -C_{in} \kappa'_{sh}, \nonumber\\
\label{freemodes}\Delta&=&-4 C_{in} C_{sh} (-K (\kappa'_{in}+\kappa'_{sh})+\kappa'_{in} \kappa'_{sh} )+Z^2. \nonumber
\end{eqnarray}
For the orbital precession given by Eq.(\ref{precession}), we have, correct up to the first order in inclinations and obliquities, the forced solution for the spin positions 
\begin{eqnarray}
S_{sh}&=&\sum_j (i_j+\varepsilon_{j,sh}) e^{I (\dot \Omega_j t+ \gamma_j-\pi/2)},\\
S_{in}&=&\sum_j (i_j+\varepsilon_{j,in}) e^{I (\dot \Omega_j t+ \gamma_j-\pi/2)},\\
\label{solution}
\varepsilon_{j,sh}&=&\frac{i_j \dot\Omega_j (K(C_{sh}+C_{in})-C_{sh} \kappa'_{in} -C_{in}\, C_{sh} \,\dot\Omega_j)}{C_{in}C_{sh}(\omega_{+}+\dot\Omega_j)(\omega_{-}+\dot\Omega_j)},\\
\label{solution2}\varepsilon_{j,in}&=&\frac{i_j \dot\Omega_j (K(C_{sh}+C_{in})-C_{in} \kappa'_{sh}-C_{in}\, C_{sh} \,\dot\Omega_j)}{C_{in}C_{sh}(\omega_{+}+\dot\Omega_j)(\omega_{-}+\dot\Omega_j)}.
\end{eqnarray}
For comparisons with the observation, we define the shell obliquity ($\varepsilon_{sh}$), its minimum and maximum values ($\varepsilon_{max,sh},\varepsilon_{min,sh}$), the deviation ($\delta_{sh}$), and two resonance factors ($fr^{\pm}_j$), which describes the amplification of $\varepsilon_{j,sh/in}$ if $\dot\Omega_j<0$ and $\omega_{\pm}+\Omega_j$ is close to zero, in the same way as for the solid case (Eqs. (\ref{obl}), (\ref{minmax}), (\ref{dev}) and (\ref{fr})).

\subsection{Numerical results}

The IAU orbit orientation used in \cite{bwstiles} is less precise than their determination of the rotation axis orientation and is the main source of error in the obliquity and deviation calculation. By comparing the IAU orbit orientation with those derived from the ephemerides of the JPL HORIZONS system and of the TASS1.6 (\cite{vienne}), we estimated the obliquity to be $\varepsilon=0.32\pm0.02^\circ$ and the deviation to be $\delta=0.12\pm0.02^\circ$.

We evaluate the solution for a representative range of hydrostatic interior structure models with homogeneous layers of constant density that are constrained by the observed mass and radius (see Table 2). The moments of inertia of the layers depend on the polar and equatorial flattenings due to rotation and static tides that are computed from the Clairaut equation, assuming hydrostatic equilibrium. Among these models, we retain those that have a moment of inertia consistent at the $3\sigma$ level with the estimated value of $0.3141\pm0.0005$ of \cite{iess}.

\begin{table}
\caption{\textbf{Size and density of the four internal layers of the Titan models consistent with the given constraints on mass and radius:} The ocean thickness and the densities of the core are calculated for the given values of the other interior parameters. Within a given set denoted by curly brackets, the values are equally spaced. }             
\label{modeles}      
\centering         
\begin{tabular}{lll}
\hline\hline
layer & Thickness(km)/Radius(km) & Density (kg\, m$^{-3}$) \\
\hline
ice shell		& $h_{sh}=\{5\},\{10, 20, ...,200\}$	& $\{800, 900,..., 1200\}$ \\
ocean			& $h_o=5-570$			& $\{1000,1100,...,1400\}$\\
ice mantle		& $R_m=\{2000, 2025, ..., 2550\}$& $\{1200,1300,1400\}$\\
ice/rock core		& $R_c=\{1400, 1410, ..., 2200\}$ & $2469-3176$ \\
\hline
\end{tabular}
\end{table}

The periods of the free  modes ($T_{\pm}=2\pi/\omega_{\pm}$) and the shell obliquity amplitudes $\varepsilon_{j,sh}$ as a function of the moment of inertia for the interior models are shown in Figs. 1 and 2, respectively. Far from the resonances ($fr_{1,2,4}\simeq 1$), the shell obliquity is below $0.15^\circ$. Therefore, a significant resonant amplification of at least one of the $\varepsilon_{j,sh}$ is needed to obtain a time-variable obliquity that can be as large as the observed one ($[\varepsilon_{max,sh};\varepsilon_{min,sh}]\cap [0.32^\circ-3 \sigma;0.32^\circ+3 \sigma]\neq \emptyset$). Some interior models have $T_{+}$ close to the opposite of the period of $\dot\Omega_1$ or $T_{-}$ very close to the opposite of the period of $\dot\Omega_4$ and have a resonant amplification of $\varepsilon_{1,sh}$ or $\varepsilon_{4,sh}$, respectively, that enables a sufficiently large obliquity value. These interior models are indicated by red and green dots in Figs. 1 and 2, which show that a specific interior model cannot be close to both resonances. Given our range of interior models (see Table 2), some models have the same moment of inertia, but not necessarily the same free mode period, and form vertical lines in Fig. 1. Because of the large number of retained interior models, these vertical lines cannot be easily distinguished, except for those appearing as peaks on the graph of $T_+$ and corresponding to models with the same $h_s, R_c, \rho_m, \rho_o$, and $\rho_s$, such as $\rho_o=\rho_m$, and with different $R_m$. One can show that $T_+$ is approximately proportional to $C_i$ for these models and is larger for larger $R_m$. The interior models at the top of the peaks (red dots), hereafter called class 1 models, are of interest because their shell obliquity can reach the observed value of the obliquity, as a result of a resonant amplification of $\varepsilon_{1,sh}$. These models are characterized by a very large interior radius of 2525-2550 km, a thin ice shell (5-30 km thickness), and a thin ocean (5-45 km thickness). Interior models that can reach the observed obliquity through a resonant amplification of $\varepsilon_{4,sh}$ (class 2 models) are more diverse than the models of the first class. They are characterized by a thicker ice shell (from 150 to 200 km), an ocean of from 5 to 425 km in thickness, an ice mantle of from 10 to 590 km in thickness and a core radius of between 1800 and 2070 km. Table 3 presents two such resonant models. Model 1 is in resonance with the main period of precession ($fr^{+}_{1}=5.31>1$) and model 2 is in a close resonance with the fourth period of precession ($fr^{-}_{4}=171.11\gg 1$).

\begin{figure}
\centering
\includegraphics[width=0.40\columnwidth,trim = 0mm 55mm 5mm 60mm,clip]{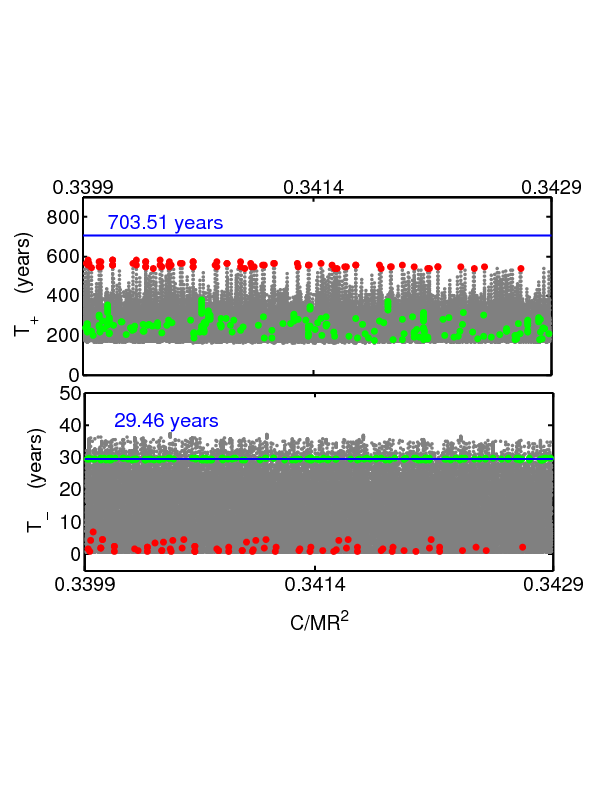}
 \caption{Periods of the 'coupled' ($T_+$) and 'decoupled' ($T_-$) free  modes as a function of the moment of inertia, for our range of interior models. The blue horizontal lines represent the opposite of the first and the fourth orbital periods. Some interior models have $T_+$ close to $703.51$ years and other interior models have a period $T_-$ very close to $29.46$ years, leading to a resonance. The red and green dots are the interior models that, thanks to a resonant amplification of $\varepsilon_{1,s}$ or $\varepsilon_{4,s}$, can reach a shell obliquity as large as the observed one (see also Fig. 2).} \label{Fig1}
 \end{figure}

\begin{figure}
\centering
\includegraphics[width=0.40\columnwidth,trim =10mm 69mm 15mm 45mm,clip]{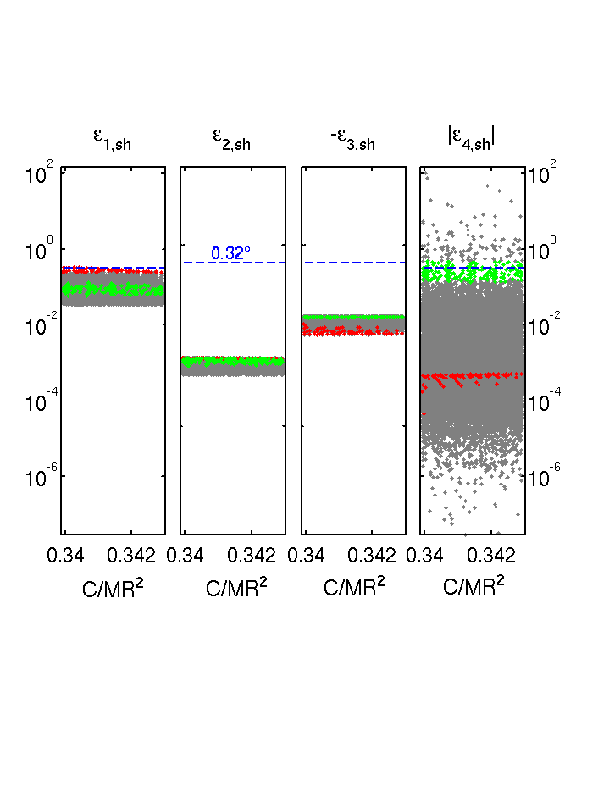}
 \caption{Shell obliquity amplitudes $\varepsilon_{j,sh}$ as a function of the moment of inertia, for our range of interior models. $\varepsilon_{1,sh}$ and $\varepsilon_{2,sh}$ are positive whereas $\varepsilon_{3,sh}$ is negative because of the positive sign of $\dot \Omega_3$. $\varepsilon_{4,sh}$ can be either negative or positive since $(\omega_{-}+\dot \Omega_4)$ can be positive or negative.  The red/green dots have the same meaning as in Fig. 1.}\label{Fig2}
 \end{figure}

\begin{table}
\caption{\textbf{Internal structure parameters, obliquities, free modes periods, and resonant factors for two resonant models:} Models 1 and 2 are in resonance with the first and the fourth period of precession, respectively.}             
\label{table3}      
\centering         
\begin{tabular}{c c c}\hline\hline
 & \textbf{Model 1 }& \textbf{Model 2 } \\
\hline
$h_{sh}$, $h_o$, $R_m$, $R_c$  		&10, 15, 2550, 1680     & 170, 105, 2300, 1890\\
$\rho_{sh}$, $\rho_o$, $\rho_m$, $\rho_c$ 	&1200, 1400, 1400, 3142 & 1100, 1400, 1400, 2758\\
\hline
$\varepsilon_{1,sh}$, $\varepsilon_{1,in}$    &\textbf{0.3198}, 1.3816	&0.0999, 0.2528 \\
$\varepsilon_{2,sh}$, $\varepsilon_{2,in}$    &0.0007, 0.0032		&0.0007, 0.0016 \\
$\varepsilon_{3,sh}$, $\varepsilon_{3,in}$    &-0.0043, -0.0126		&-0.0102, -0.0125 \\
$\varepsilon_{4,sh}$, $\varepsilon_{4,in}$    &-0.0004, -0.0023		&\textbf{0.2300}, -0.0270 \\
$T_{+}$, $T_{-}$   &572.019, 2.342 &300.391, 29.285 \\
\hline
$fr^{rig}_{1}$, $fr^{rig}_{2}$, $fr^{rig}_{4}$    &\textbf{5.31}, 1.21, 1.05&1.75, 1.10, 1.11\\
$fr^{dec}_{1}$, $fr^{dec}_{2}$, $fr^{dec}_{4}$    &1.00, 1.00, 1.09& 1.04, 1.01, \textbf{171.11}\\
\end{tabular}
\end{table}

\begin{figure}
\centering
\includegraphics[width=0.40\columnwidth]{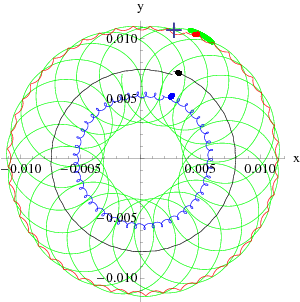}
\caption{Projection of the normal to the orbit (blue) and of the rotation axis on the Laplace plane (black for the solid case, red for model 1 and green for model 2) over the period of the main precession, beginning at J2000 (thin curves) and over the observation period of \cite{bwstiles} (thick curves). The "+" marker is the projection of the rotation axis measured by \cite{bwstiles}. (Unit of the graph is radian.)}\label{Fig3}
\end{figure}

In addition to the obliquity, the deviation of the shell rotation axis from the plane defined by the orbit and Laplace plane normals (see Eq.(\ref{dev})) has to be consistent with the observations. With a small deviation of $0.03^\circ$ at best, the class 1 models cannot explain the observed deviation of about $0.12^\circ$. The deviation reported by \cite{bwstiles} can be seen as an averaged position of the rotation axis during the period of coverage of the analysed radar observations (from Oct 26 2004 to Feb 22 2007), during which model 2 has a deviation smaller than $0.12^\circ$.
This is true for all the models of class 2 with a positive $\varepsilon_{4,sh}$, because the deviation depends mainly on the ephemerides, particularly on the phases of the orbital precession. We would need to add about $15^\circ$ and $45^\circ$ to $\gamma_4$ to obtain the correct deviation at time Oct 26 2004 and Feb 22 2007, respectively. However, a comparison of TASS1.6 with a series expansion fitted on the ephemerides of the JPL HORIZONS system convinced us that the phase $\gamma_4$ in TASS1.6 is correct up to a few degrees. The series expansion given in \cite{vienne} is incomplete and additional terms with smaller amplitudes exist (up to $j=21$, \cite{vienneperscom}). With those additional terms, some internal models that are consistent with the moment of inertia can account for both the estimated obliquity and deviation, thanks to a combination of a very close resonance between $\omega_+$ and $\dot \Omega_{14}=-2\pi/578.73 y$, besides the further resonance with $\dot \Omega_1$, discussed before. However, this very close resonance seems implausible because of the small fraction of suitable internal structure models.
\begin{figure}
\centering
\includegraphics[width=0.40\columnwidth,trim = 4mm 0mm 4mm 0mm,clip]{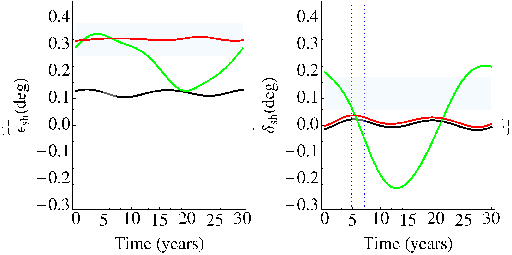}\\
\caption{Shell obliquity $\varepsilon_{sh}$ (left) and deviation $\delta_{sh}$ (right) over 30 years beginning on J2000 (black for the solid case, red for model 1 and green for model 2). The vertical blue lines show the observation period of \cite{bwstiles} and the blue boxes represent our estimated obliquity ($0.32^\circ \pm 3 \sigma$) and deviation ($0.12^\circ \pm 3 \sigma$).} \label{Fig4}
\end{figure}

\section{Discussion and perspectives} 

Since we have found a better consistency between the measured and computed positions of the rotation axis with our model than in the solid case, assuming that Titan is locked in the Cassini state, our study is an indication of a possible water ocean layer beneath the surface of Titan.

For two classes of interior models with a liquid water ocean beneath an ice shell, the obliquity computed with our new Cassini state model can be as large as the observed one. However, we rejected the first class of models with a very thin ice shell and a shallow liquid ocean since a thin shell implies high temperatures and therefore, according to the water phase diagram, a thick ocean.
On the other hand, models of class 2 are possible but can be quite different so that the estimated obliquity and moment of inertia do not provide accurate constraints on the interior structure.

The theoretically predicted deviation of the rotation axis with respect to the plane defined by the orbit and Laplace plane normals is inconsistent with the observations. We have shown that the use of ephemerides extended to additional frequencies could solve this problem, but only for a small fraction of the rejected class 1 models. Another explanation is that Titan may be slightly offset from the Cassini state because of a recent excitation. However, other explanations extending our Cassini state model might be tested. We have neglected the viscosity at the ocean boundaries because we have assumed that the timescale on which the viscosity is effective is greater than the timescale of the node precession of Titan. We have also neglected the polar motion induced by the atmosphere and the resulting variations in the rotation axis orientation in space since they are expected to have a very small amplitude compared to Titan's obliquity.
Although the deformation of the ice shell has not been included, our new Cassini state model is a first step towards a more realistic model of the obliquity of an icy satellite with an ocean.

\section*{acknowledgements}
    R.-M. Baland is a research fellow of the Fonds pour la formation \`{a} la Recherche dans l'Industrie et dans l'Agriculture (FRIA), Belgium. This work was financially supported by the Belgian PRODEX program managed by the European Space Agency in collaboration with the Belgian Federal Science Policy Office.

\bibliographystyle{abbrv}

\bibliography{articlearxive} 

\begin{thebibliography}{10}

\bibitem{baland}
R.-M. {Baland} and T.~{Van Hoolst}.
\newblock {Librations of the Galilean satellites: The influence of global
  internal liquid layers}.
\newblock {\em Icarus}, 2010.

\bibitem{billsepsc09}
B.~{Bills} and F.~{Nimmo}.
\newblock {Spin, gravity, and moments of inertia of Titan}.
\newblock In {\em European Planetary Science Congress 2009}, pages 553--+,
  Sept. 2009.

\bibitem{bills05}
B.~G. {Bills}.
\newblock {Free and forced obliquities of the Galilean satellites of Jupiter}.
\newblock {\em Icarus}, 175:233--247, May 2005.

\bibitem{iess}
L.~{Iess}, N.~J. {Rappaport}, R.~A. {Jacobson}, P.~{Racioppa}, D.~J.
  {Stevenson}, P.~{Tortora}, J.~W. {Armstrong}, and S.~W. {Asmar}.
\newblock {Gravity Field, Shape, and Moment of Inertia of Titan}.
\newblock {\em Science}, 327:1367--, Mar. 2010.

\bibitem{seidelmann}
P.~K. {Seidelmann}, B.~A. {Archinal}, M.~F. {A'Hearn}, D.~P. {Cruikshank},
  J.~L. {Hilton}, H.~U. {Keller}, R.~J. {Oberst}, J.~L. {Simon}, P.~{Stooke},
  D.~J. {Tholen}, and P.~C. {Thomas}.
\newblock {Working Group on Cartographic Coordinates and Rotational Elements}.
\newblock {\em Transactions of the International Astronomical Union, Series A},
  26:181--181, Mar. 2007.

\bibitem{bwstiles}
B.~W. {Stiles}, R.~L. {Kirk}, R.~D. {Lorenz}, S.~{Hensley}, E.~{Lee}, S.~J.
  {Ostro}, M.~D. {Allison}, P.~S. {Callahan}, Y.~{Gim}, L.~{Iess}, P.~{Perci
  del Marmo}, G.~{Hamilton}, W.~T.~K. {Johnson}, R.~D. {West}, and {The Cassini
  RADAR Team}.
\newblock {Determining Titan's Spin State from Cassini RADAR Images}.
\newblock {\em The Astronomical Journal}, 135:1669--1680, May 2008.

\bibitem{bwstiles2}
B.~W. {Stiles}, R.~L. {Kirk}, R.~D. {Lorenz}, S.~{Hensley}, E.~{Lee}, S.~J.
  {Ostro}, M.~D. {Allison}, P.~S. {Callahan}, Y.~{Gim}, L.~{Iess}, P.~{Perci
  del Marmo}, G.~{Hamilton}, W.~T.~K. {Johnson}, R.~D. {West}, and {The Cassini
  RADAR Team}.
\newblock {ERRATUM: ''Determining Titan's Spin State from Cassini Radar
  Images'' <A href=''bib\_query$\backslash$?2008AJ....135.1669S''>(2008, AJ,
  135, 1669)</A>}.
\newblock {\em The Astronomical Journal}, 139:311--+, Jan. 2010.

\bibitem{zseto}
A.~M.~K. {Szeto} and S.~{Xu}.
\newblock {Gravitational coupling in a triaxial ellipsoidal Earth}.
\newblock {\em Journal of Geophysical Research}, 102:27651--27658, Dec. 1997.

\bibitem{vanhoolst}
T.~{Van Hoolst}, N.~{Rambaux}, {\"O}.~{Karatekin}, and R.~{Baland}.
\newblock {The effect of gravitational and pressure torques on Titan's
  length-of-day variations}.
\newblock {\em Icarus}, 200:256--264, Mar. 2009.

\bibitem{vienneperscom}
A.~{Vienne}.
\newblock {Personal communication}.
\newblock {\em Personal communication}, 2010.

\bibitem{vienne}
A.~{Vienne} and L.~{Duriez}.
\newblock {TASS1.6: Ephemerides of the major Saturnian satellites.}
\newblock {\em Astronomy And Astrophysics}, 297:588--+, May 1995.

\end{thebibliography}
\nocite{seidelmann}

\end{document}